\documentclass[prl,amsfonts,amssymb,floats,twocolumn,superscriptaddress,aps,floatfix]{revtex4-2}
\renewcommand{\paragraph}[1]{\textit{#1.}\textemdash}
\newcommand{\ui}{\mathrm{i}}

\makeatletter
\renewcommand{\fnum@figure}{\hspace{11pt}FIG. \thefigure}
\makeatother

\usepackage{comment} % enables the use of multi-line comments (\ifx \fi) 
\usepackage{graphicx}
\usepackage{pythontex}
\usepackage{bm}
\usepackage{amsmath}
\usepackage[utf8]{inputenc}
\usepackage{amssymb}
\usepackage{dsfont}
\usepackage{mathrsfs}
\usepackage[english]{babel} 
\usepackage{enumitem}
\usepackage{braket} % brakets with \bra{..} and \ket{..}
\usepackage{bbold}  % identity with \mathbb{1}
\usepackage{listings}
\usepackage{booktabs} % for prettier tables
\usepackage[colorlinks=true, breaklinks=true, linkcolor=blue, citecolor=blue, urlcolor=blue]{hyperref} 
\usepackage{MnSymbol}
\usepackage{orcidlink}
\usepackage{float}
\usepackage{nicefrac}
\usepackage{xfrac}

\renewcommand{\k}{{\ensuremath{\mathbf{k}}}}

\renewcommand{\r}{{\ensuremath{\mathbf{r}}}}

\usepackage{xcolor}

\newcommand{\locS}{\mathcal{S}}
\newcommand{\locSb}{\bm{\mathcal{S}}}
\renewcommand{\vr}{\mathbf{r}}

\newcommand{\vd}{\bm{\delta}}
\newcommand{\kb}{\mathbf{k}}

\newcommand{\cre}{\hat{a}^\dagger}
\newcommand{\ann}{\hat{a}^{\phantom{\dagger}}}

\newcommand{\annB}{\hat{b}^{\phantom{\dagger}}}

\newcommand{\sumprime}{\sideset{}{'}\sum}
\newcommand{\panel}[1]{{\color{blue}\textbf{#1}}}
\newcommand{\panelk}[1]{{\color{black}\textbf{#1}}}

\makeatletter
\newcommand{\appendixtableofcontents}{%
  \section*{Appendix Contents}%
  \@starttoc{atoc}%
}

\newcommand{\appsection}[1]{%
  \section{#1}%
  \addcontentsline{atoc}{section}{\protect\numberline{\thesection}#1}%
}
\newcommand{\appsubsection}[1]{%
  \subsection{#1}%
  \addcontentsline{atoc}{subsection}{\protect\numberline{\thesubsection}#1}%
}

\makeatother

\newcommand{\LatticeTetrahedral}{
\begin{tikzpicture}[scale=0.95, line cap=round, line join=round]

\def\Nx{6} % number of sites in a1-direction
\def\Ny{6} % number of sites in a2-direction
\def\a{0.7} % lattice spacing

\pgfmathsetmacro{\sx}{0.5*\a}
\pgfmathsetmacro{\sy}{0.866025403784*\a} % sqrt(3)/2 * a

\definecolor{colA}{HTML}{FAD961}   % red #FAD961
\definecolor{colB}{HTML}{4820FA}  % blue #4820FA
\definecolor{colC}{HTML}{53DAFA}   % green #53DAFA
\definecolor{colD}{HTML}{FA3774}  % purple #FA3774

\foreach \i in {0,...,\numexpr\Nx-1\relax} {
  \foreach \j in {0,...,\numexpr\Ny-1\relax} {

    \coordinate (P) at ({\a*\i + \sx*\j},{\sy*\j});

    \ifnum\i<\numexpr\Nx-1\relax
      \coordinate (Pr) at ({\a*(\i+1) + \sx*\j},{\sy*\j});
      \draw[black!12, line width=0.6pt] (P) -- (Pr);
    \fi

    \ifnum\j<\numexpr\Ny-1\relax
      \coordinate (Pu) at ({\a*\i + \sx*(\j+1)},{\sy*(\j+1)});
      \draw[black!12, line width=0.6pt] (P) -- (Pu);
    \fi

    \ifnum\i<\numexpr\Nx-1\relax
      \ifnum\j>0\relax
        \coordinate (Pdiag) at ({\a*(\i+1) + \sx*(\j-1)},{\sy*(\j-1)});
        \draw[black!12, line width=0.6pt] (P) -- (Pdiag);
      \fi
    \fi
  }
}

\foreach \i in {0,...,\numexpr\Nx-1\relax} {
  \foreach \j in {0,...,\numexpr\Ny-1\relax} {

    \coordinate (P) at ({\a*\i + \sx*\j},{\sy*\j});

    \pgfmathtruncatemacro{\imod}{mod(\i,2)}
    \pgfmathtruncatemacro{\jmod}{mod(\j,2)}
    \pgfmathtruncatemacro{\s}{2*\imod + \jmod}

    \ifnum\s=0
      \def\thiscol{colA}
    \fi
    \ifnum\s=1
      \def\thiscol{colB}
    \fi
    \ifnum\s=2
      \def\thiscol{colC}
    \fi
    \ifnum\s=3
      \def\thiscol{colD}
    \fi

    \filldraw[fill=\thiscol, draw=black!55, line width=0.5pt] (P) circle (0.1);

  }
}

\def\offx{-1}
\def\offy{-1.1}
\begin{scope}[xshift={(\Nx+1.2)*\a}, yshift={(\Ny-1)*\sy}]
  \node[anchor=west] at (-1.8+\offx,4.85+\offy) {\small \bfseries Tetrahedral Order. Configuration 1};
  \filldraw[fill=colA, draw=black!55, line width=0.5pt] (-0.1+\offx,4.10+\offy) circle (0.1);
  \node[anchor=west] at (0.25+\offx,4.10+\offy) {\small A};
  \filldraw[fill=colB, draw=black!55, line width=0.5pt] (-0.1+\offx,4-0.35+\offy) circle (0.1);
  \node[anchor=west] at (0.25+\offx,4-0.35+\offy) {\small B};
  \filldraw[fill=colC, draw=black!55, line width=0.5pt] (-0.1+\offx,4-0.80+\offy) circle (0.1);
  \node[anchor=west] at (0.25+\offx,4-0.80+\offy) {\small C};
  \filldraw[fill=colD, draw=black!55, line width=0.5pt] (-0.1+\offx,4-1.25+\offy) circle (0.1);
  \node[anchor=west] at (0.25+\offx,4-1.25+\offy) {\small D};
\end{scope}

\end{tikzpicture}
}

\newcommand{\LatticeTetrahedralTwo}{
\begin{tikzpicture}[scale=0.95, line cap=round, line join=round]

\def\Nx{4} % number of sites in a1-direction
\def\Ny{4} % number of sites in a2-direction
\def\a{0.7} % lattice spacing

\pgfmathsetmacro{\sx}{0.5*\a}
\pgfmathsetmacro{\sy}{0.866025403784*\a} % sqrt(3)/2 * a

\definecolor{colA}{HTML}{FAD961}   % red #FAD961
\definecolor{colB}{HTML}{4820FA}  % blue #4820FA
\definecolor{colC}{HTML}{53DAFA}   % green #53DAFA
\definecolor{colD}{HTML}{FA3774}  % purple #FA3774

\foreach \i in {0,...,\numexpr\Nx-1\relax} {
  \foreach \j in {0,...,\numexpr\Ny-1\relax} {

    \coordinate (P) at ({\a*\i + \sx*\j},{\sy*\j});

    \ifnum\i<\numexpr\Nx-1\relax
      \coordinate (Pr) at ({\a*(\i+1) + \sx*\j},{\sy*\j});
      \draw[black!12, line width=0.6pt] (P) -- (Pr);
    \fi

    \ifnum\j<\numexpr\Ny-1\relax
      \coordinate (Pu) at ({\a*\i + \sx*(\j+1)},{\sy*(\j+1)});
      \draw[black!12, line width=0.6pt] (P) -- (Pu);
    \fi

    \ifnum\i<\numexpr\Nx-1\relax
      \ifnum\j>0\relax
        \coordinate (Pdiag) at ({\a*(\i+1) + \sx*(\j-1)},{\sy*(\j-1)});
        \draw[black!12, line width=0.6pt] (P) -- (Pdiag);
      \fi
    \fi
  }
}

\foreach \i in {0,...,\numexpr\Nx-1\relax} {
  \foreach \j in {0,...,\numexpr\Ny-1\relax} {

    \coordinate (P) at ({\a*\i + \sx*\j},{\sy*\j});

    \pgfmathtruncatemacro{\imod}{mod(\i,2)}
    \pgfmathtruncatemacro{\jmod}{mod(\j,2)}
    \pgfmathtruncatemacro{\s}{2*\imod + \jmod}

    \ifnum\s=0
      \def\thiscol{colB}
    \fi
    \ifnum\s=1
      \def\thiscol{colC}
    \fi
    \ifnum\s=2
      \def\thiscol{colA}
    \fi
    \ifnum\s=3
      \def\thiscol{colD}
    \fi

    \filldraw[fill=\thiscol, draw=black!55, line width=0.5pt] (P) circle (0.1);

  }
}

\begin{scope}[xshift={(\Nx+1.2)*\a}, yshift={(\Ny-1)*\sy}]
  \node[anchor=west] at (-0.4,2.35) {\small \bfseries Configuration 2};
\end{scope}

\end{tikzpicture}
}

\newcommand{\LatticeTetrahedralThree}{
\begin{tikzpicture}[scale=0.95, line cap=round, line join=round]

\def\Nx{4} % number of sites in a1-direction
\def\Ny{4} % number of sites in a2-direction
\def\a{0.7} % lattice spacing

\pgfmathsetmacro{\sx}{0.5*\a}
\pgfmathsetmacro{\sy}{0.866025403784*\a} % sqrt(3)/2 * a

\definecolor{colA}{HTML}{FAD961}   % red #FAD961
\definecolor{colB}{HTML}{4820FA}  % blue #4820FA
\definecolor{colC}{HTML}{53DAFA}   % green #53DAFA
\definecolor{colD}{HTML}{FA3774}  % purple #FA3774

\foreach \i in {0,...,\numexpr\Nx-1\relax} {
  \foreach \j in {0,...,\numexpr\Ny-1\relax} {

    \coordinate (P) at ({\a*\i + \sx*\j},{\sy*\j});

    \ifnum\i<\numexpr\Nx-1\relax
      \coordinate (Pr) at ({\a*(\i+1) + \sx*\j},{\sy*\j});
      \draw[black!12, line width=0.6pt] (P) -- (Pr);
    \fi

    \ifnum\j<\numexpr\Ny-1\relax
      \coordinate (Pu) at ({\a*\i + \sx*(\j+1)},{\sy*(\j+1)});
      \draw[black!12, line width=0.6pt] (P) -- (Pu);
    \fi

    \ifnum\i<\numexpr\Nx-1\relax
      \ifnum\j>0\relax
        \coordinate (Pdiag) at ({\a*(\i+1) + \sx*(\j-1)},{\sy*(\j-1)});
        \draw[black!12, line width=0.6pt] (P) -- (Pdiag);
      \fi
    \fi
  }
}

\foreach \i in {0,...,\numexpr\Nx-1\relax} {
  \foreach \j in {0,...,\numexpr\Ny-1\relax} {

    \coordinate (P) at ({\a*\i + \sx*\j},{\sy*\j});

    \pgfmathtruncatemacro{\imod}{mod(\i,2)}
    \pgfmathtruncatemacro{\jmod}{mod(\j,2)}
    \pgfmathtruncatemacro{\s}{2*\imod + \jmod}

    \ifnum\s=0
      \def\thiscol{colB}
    \fi
    \ifnum\s=1
      \def\thiscol{colD}
    \fi
    \ifnum\s=2
      \def\thiscol{colC}
    \fi
    \ifnum\s=3
      \def\thiscol{colA}
    \fi

    \filldraw[fill=\thiscol, draw=black!55, line width=0.5pt] (P) circle (0.1);

  }
}

\begin{scope}[xshift={(\Nx+1.2)*\a}, yshift={(\Ny-1)*\sy}]
  \node[anchor=west] at (-0.4,2.35) {\small \bfseries Configuration 3};
\end{scope}

\end{tikzpicture}
}

\begin{document}

\title{Dynamical dimer structure factor of the triangular $S=\nicefrac{1}{2}$ Heisenberg antiferromagnet}

\author{Markus Drescher \orcidlink{0000-0003-1293-2844}}
\thanks{markus.drescher@tum.de}
\affiliation{Department of Physics, Technische Universität München, 85748 Garching, Germany}%

\author{Laurens Vanderstraeten \orcidlink{0000-0002-3227-9822}}
\affiliation{Center for Nonlinear Phenomena and Complex Systems, Université Libre de Bruxelles, 1050 Brussels, Belgium}%

\author{Roderich Moessner}
\affiliation{Max-Planck-Institut für Physik komplexer Systeme, 01187 Dresden, Germany}%

\author{Frank Pollmann \orcidlink{0000-0003-0320-9304}}
\affiliation{Department of Physics, Technische Universität München, 85748 Garching, Germany}%
\affiliation{Munich Center for Quantum Science and Technology (MCQST), 80799 Munich, Germany}%

\author{Johannes Knolle \orcidlink{0000-0002-0956-2419}}
\affiliation{Department of Physics, Technische Universität München, 85748 Garching, Germany}%
\affiliation{Munich Center for Quantum Science and Technology (MCQST), 80799 Munich, Germany}%

\date{\today}

\begin{abstract}
The dynamical dimer structure factor is an observable probing spin-singlet excitations of quantum magnets distinct from those commonly studied by the spin structure factor. We report the dimer response for the extended spin-$1/2$ antiferromagnetic Heisenberg model on the triangular lattice using large-scale GPU-accelerated matrix-product-state simulations.
We investigate the ordered phases with $120^\circ$ coplanar, collinear stripe, and tetrahedral spin order, as well as candidate quantum spin-liquid (QSL) regimes, comprising an expected gapless $U(1)$ Dirac QSL  and a chiral QSL at finite spin-scalar-chirality coupling.
In the ordered phases, we find low-energy modes below the onset of the two-magnon continuum illustrating avoided quasiparticle decay.
Within the candidate gapless QSL, we observe absolute dispersion minima at momenta of half the Brillouin zone corners, $X\equiv K/2$, in agreement with field-theory predictions that singlet monopole excitations of the $U(1)$ Dirac spin liquid become gapless at these points. Thus, the high-resolution dynamical dimer response provides support for a $U(1)$ Dirac QSL with singlet monopole excitations.
\end{abstract}

\maketitle

\paragraph{Introduction}%
The triangular-lattice Heisenberg antiferromagnet constitutes a paradigmatic model of frustrated quantum magnetism, extensively studied in both theory and experiment~\cite{frustratedmagnetism}.
In spin-\nicefrac{1}{2} systems, the interplay of geometric frustration and strong quantum fluctuations makes it a promising platform for realizing quantum spin-liquid (QSL) phases~\cite{Anderson1973, IoffeLarkin1989, Wen1991_topologicalorder, ReadSachdev1991, SenthilFisher2000, Wen2002, Hermele2004, Coldea2001, Balents2010, Knolle2019, Broholm2020}.
Despite advances in candidate materials and experimental neutron scattering techniques~\cite{Ito2017, Ranjith2019, Dai2021, Zhang2021, Scheie2024, Cao2025}, 
resolving the spin excitation spectrum remains challenging both experimentally and numerically~\cite{Gohlke2017, Verresen2018, Ferrari2019, Mourigal2013, Ghioldi2018, Sherman2023, Drescher2023, Bag2024, scheie_battista_2024}.
This complicates the interpretation of the widely studied dynamical spin structure factor and limits insight into the nature of the ground state and its excitations~\cite{Ferrari2021}.

It is therefore desirable to explore alternative observables such as the dimer (i.e., $\mathrm{SU}(2)$-singlet) correlations that probe additionally the spin-singlet sector, whose constitutive importance was emphasized by Anderson in one of the founding works of quantum spin-liquid physics \cite{Anderson1973}. Indeed, the Rokhsar-Kivelson quantum dimer model was formulated precisely to capture the physics of singlet-dominated phases~\cite{RokhsarKivelson1988}, and it was in this model on the triangular lattice that a fractionalized $\mathbb{Z}_2$ quantum spin-liquid phase---the resonating valence-bond liquid---was first discovered~\cite{Moessner2001}. In addition, dimer correlations may become experimentally accessible through Raman spectroscopy or resonant inelastic X-ray scattering (RIXS)~\cite{ament2011resonant,LiJinDattaYao2023, Luo2015,KimChaloupkaSinghKimKimCasaSaidHuangGog2020,Xiong2020,MitranoJohnstonKimDean2024}. Only few works explored their dynamical signatures in frustrated spin systems~\cite{Lu2026}. For example, employing the exact solution of the honeycomb Kitaev QSL, they directly probe dynamical Majorana excitations~\cite{knolle2014raman,halasz2016resonant,halasz2019observing} without coupling to fluxes, which dominate and complicate the interpretation of the spin structure factor~\cite{knolle2014dynamics}. However, the phenomenology of the dynamical dimer response and possible coupling to other types of exotic fractionalized excitations are missing for strongly coupled QSL phases calling for state-of-the-art numerical calculations.

\begin{figure}
    \centering
    \includegraphics[scale=1]{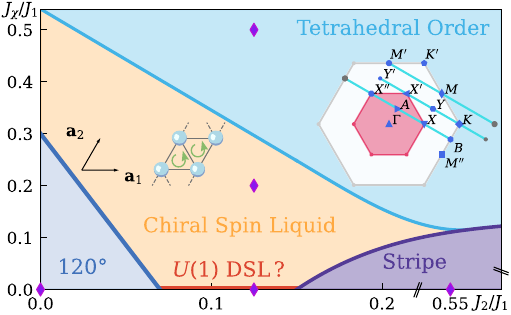}
    \caption{Schematic phase diagram of the extended Heisenberg model on the triangular lattice with a scalar chiral interaction. The qualitative phase boundaries follow previous numerical studies (cf. Ref.~\cite{Wietek2017, Gong2017}). The inset within the chiral spin-liquid regime shows the triangular lattice with Bravais vectors $\mathbf{a}_1 = (1,0)^T$ and $\mathbf{a}_2 = (1/2, \sqrt{3}/2)^T$. The chirality of the $J_\chi$ term is uniform across all triangular plaquettes (green arrows). The Brillouin zone on the right indicates the momentum cuts accessible on the $\mathrm{YC}6$ cylinder, for which the dynamical structure factor is computed in this work. The violet diamonds mark the points in parameter space where the dimer structure factor is evaluated.}
    \label{fig:schematic_phase_diagram}
\end{figure}

In this work, we study the dynamical dimer structure factor of the extended triangular-lattice Heisenberg model including next-nearest-neighbor exchange and a scalar chiral interaction~\cite{Wen1989, Baskaran1989, Wietek2017, Gong2017}. The Hamiltonian reads~\cite{Bauer2014, Wietek2017, Gong2017}
\begin{align}
    \hat{H} &= J_1 \sum_{\langle i,j\rangle} \hat{\mathbf{S}}_i \cdot \hat{\mathbf{S}}_j
    + J_2 \sum_{\llangle i,j\rrangle} \hat{\mathbf{S}}_i \cdot \hat{\mathbf{S}}_j \nonumber\\
    &\phantom{=}+ J_\chi \sum_{i,j,k\in \triangle, \triangledown} \hat{\mathbf{S}}_i \cdot \left(\hat{\mathbf{S}}_j \times \hat{\mathbf{S}}_k\right)\,,
    \label{equ:hamiltonian}
\end{align}
where $J_1$ and $J_2$ denote the nearest- and next-nearest-neighbor Heisenberg exchange couplings and the chiral term sums over triangular plaquettes with a fixed site ordering to preserve local spin chirality (cf. Fig.~\ref{fig:schematic_phase_diagram}).

The phase diagram of this model is illustrated schematically in Fig.~\ref{fig:schematic_phase_diagram}\panel{a} following Refs.~\cite{Wietek2017, Gong2017}.
Against Anderson's proposal of a resonating valence bond state~\cite{Anderson1973, Fazekas1974}, semiclassical and numerical investigations revealed a coplanar $120^\circ$ spin order at the Heisenberg point~\cite{Jolicoeur1990, Chubukov1994, Bernu1994, Capriotti1999, White2007, Chernyshev2006, Starykh2006, Chernyshev2009}. This magnetic order extends to a finite region in the augmented parameter space~\cite{Wietek2017, Gong2017}.
At $J_\chi = 0$, increasing frustration is widely believed to drive the system into a disordered regime for intermediate couplings $0.07\lesssim J_2/J_1 \lesssim 0.15$, embedded between the $120^\circ$ and stripe orders~\cite{Kaneko2014, Zhu2015, Hu2015, Iqbal2016}.
The nature of this phase has been intensely debated, with proposals ranging from a gapped $\mathbb{Z}_2$ spin liquid~\cite{Wang2006, Zhu2015, Hu2015, JiangWhite2026} to a gapless $U(1)$ Dirac spin liquid (DSL) described by $\mathrm{QED}_3$ with $N=4$ Dirac spinons~\cite{Hermele2004, Hermele2005, Iqbal2016, Hu2019, Ferrari2019, Wietek2024}.

Upon turning on the chiral interaction $J_\chi$, the coplanar and collinear orders remain stable over extended regions.
The candidate QSL phase, however, has been shown to transition already for small couplings $J_\chi$ into a chiral spin liquid (CSL) phase~\cite{Wietek2017, Gong2017},
a topologically ordered state originally proposed by Wen, Wilczek and Zee~\cite{Wen1989} following the seminal contribution by Kalmeyer and Laughlin~\cite{Kalmeyer1987, Kalmeyer1989}.
It occurs around the classical phase transition line between the $120^\circ$ order and a tetrahedral spin order~\cite{Messio2011, Gong2017}. The latter is stabilized at sufficiently large chiral interactions.

Using large-scale matrix-product-state (MPS) simulations~\cite{Schollwoeck2011} accelerated by graphics processing units (GPUs)~\cite{li_numerical_2020, Pan2022, unfried_fast_2023,Drescher2025}, we compute the time-resolved dimer correlations in all ordered phases as well as in the candidate spin-liquid regimes.
The findings in the $120^\circ$ and tetrahedral phases underline, despite the unambiguous identification of the spectral dimer features from linear spin-wave theory (LSWT), the importance of magnon-interaction effects beyond spin-wave analysis: the phenomenon of level repulsion leads to stable low-energy modes through the Brillouin zone that manifest also as essential signatures of the dimer structure factor.

Most importantly, the spectral signatures of the candidate QSL regime at intermediate couplings $0.07\lesssim J_2/J_1 \lesssim 0.15$ reveal clear distinctions from the adjacent $120^\circ$ ordered phase. This becomes evident from the deviating behavior at the corners of the Brillouin zone: while the $120^\circ$ phase exhibits dispersion minima at $K$, the paramagnetic regime displays maxima.
This finding appears to rule out a gapped $\mathbb{Z}_2$ QSL interpretation of the low-energy sector on the $\mathrm{YC}6$ cylinder (cf. Ref.~\cite{JiangWhite2026}).
Moreover, the dimer spectrum of the $J_1-J_2$ QSL shows its global minimum at the $X=K/2$ points.
This observation aligns with the field-theoretical predictions of gapless singlet monopole excitations at these momenta, i.e., instanton events inserting a global $2\pi$ flux into the gauge background~\cite{Song2019, Song2020}.
Recent works probing static~\cite{seifert2024spin} and dynamical~\cite{FerrariWillsher2025}  spin-lattice distortions of the QSL regime on the triangular lattice have been interpreted as signatures of critical spin-singlet monopole correlations in the ground state. The dynamical dimer response reported here provides further evidence for the presence of singlet monopole excitations, supporting the interpretation of the underlying ground state being a $U(1)$ Dirac QSL.

\begin{figure*}
    \centering
    \includegraphics[scale=1]{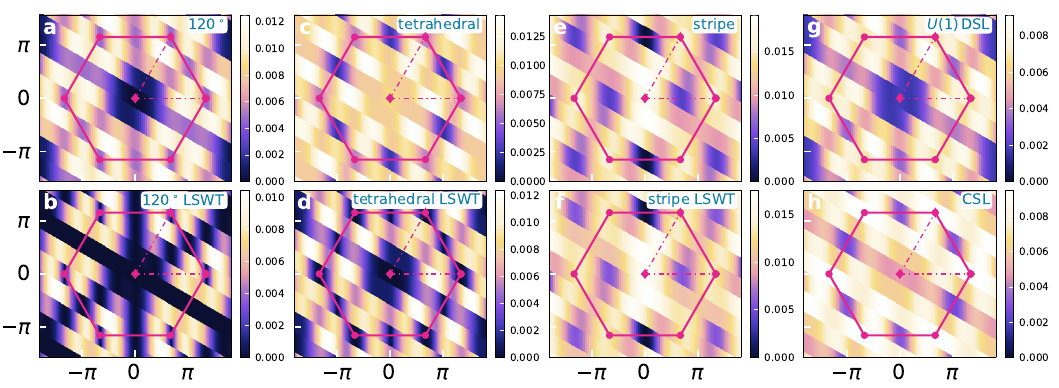}
    \caption{The static dimer structure factor $\chi_D(\kb)$ for the various ordered and liquid phases considered in this work. 
    The left column shows the results for the $120^\circ$ phase at $J_2 = 0$ from MPS (top) and LSWT (bottom), the following column for the tetrahedral phase for $J_2/J_1 =0.125$ and $J_\chi/J_1 = 0.5$ (MPS, \panelk{c}, LSWT, \panelk{d}), the third column for the stripe order with $J_2/J_1 = 0.55$ (MPS, \panelk{e}, spin-wave result, \panelk{f}). The last column shows the numerical data for the candidate $J_1-J_2$ QSL (\panelk{g}) and the chiral spin liquid (\panelk{h}). The former ground state at $J_2/J_1 = 0.125$ ($J_\chi = 0$) was stabilized in the odd (low-energy) sector (cf. End Matter), the ground state in the CSL for $J_2/J_1 =0.125$ together with $J_\chi / J_1 = 0.2$ represents the trivial topological sector. All data are for $\mathrm{YC}6$ cylinders.
    The MPS bond dimension is $\chi_{\mathrm{max}}=2000$. The cylinder lengths in the numerical simulations are $L_x = 51$ ($120^\circ$ order and $J_1-J_2$ QSL), $L_x = 81$ (CSL and tetrahedral order) and $L_x = 126$ (stripe phase). Profile plots along the individual cuts accessible in the $\mathrm{YC}6$ cylinder geometry can be found in the Supplemental Material~\cite{supp}.
    }
    \label{fig:dimer_ssf_all_phases}
\end{figure*}

\paragraph{Numerical protocol}%
Whereas the MPS methodology has been applied in different works~\cite{Verresen2019, Sherman2023, Drescher2023, Xie2023, JiangWhite2026, KovalskaDelft2026} in the last years for estimating the spin dynamical structure factor (DSF), and has become an invaluable tool to support experiments in inelastic neutron scattering, here we apply it for the first time to the more demanding case of resolving the dimer DSF.

The spin bilinear $\hat{\mathbf{S}}_i\cdot \hat{\mathbf{S}}_j$, which appears in the projector onto the spin-singlet component of a bond $(i,j)$, forms the fundamental building block of the Loudon-Fleury operator describing Raman scattering in the nearest-neighbor Heisenberg model (up to polarization-dependent prefactors)~\cite{FleuryLoudon1968, ElliottThorpe1969, ShastryShraiman1990, PerkinsBrenig2008, Wulferding2019}, as well as of the indirect non-spin-flip RIXS channel~\cite{ament2011resonant}.
As a related quantity, capturing the spin-singlet fluctuations, we introduce the dimer operator around a central site~$i$ defined as
\begin{equation}
    \hat{\mathcal{D}}_i = \frac{1}{N_\delta} \sum_\delta \left(\hat{\mathbf{S}}_i \cdot \hat{\mathbf{S}}_{i+\delta} - \braket{\hat{\mathbf{S}}_i \cdot \hat{\mathbf{S}}_{i+\delta}}\right).
    \label{equ:dimer_operator}
\end{equation}
The sum runs over all nearest-neighbor sites of $i$ labeled by $\delta$ with $N_\delta = 6$ being the coordination number of the lattice. Note that the subtraction of the dimer expectation value, which is non-vanishing, ensures that the dimer operator probes the excitation spectrum of the ground state via fluctuations around it.
To this purpose, we have to compute the connected time-resolved dimer-dimer correlation function
\begin{equation}
    C_D^{\mathrm{conn.}}(i,j;t) = \braket{\hat{\mathcal{D}}_i(t)\hat{\mathcal{D}}_j(0)}\,.
\end{equation}
The static dimer structure factor is straight-forwardly obtained from the equal-time dimer correlations:
\begin{equation}
    \chi_D(\k) = \frac{1}{N} \sum_{i,j} e^{-\ui \k \cdot(\r_i - \r_j)} \braket{\hat{\mathcal{D}}_i \hat{\mathcal{D}}_j}\,,
\end{equation}
where $N$ denotes the number of lattice sites.
Exploiting the translational invariance of the system, we obtain the dynamical dimer structure factor via
\begin{equation}
    S_D(\k,\omega) = \int \mathrm{d}t \sum_j e^{\ui \omega t - \ui \k \cdot (\r_j-\r_{j_c})}
    \braket{\hat{\mathcal{D}}_j(t) \hat{\mathcal{D}}_{j_c}(0)}\,.
\end{equation}

We first use the density renormalization group algorithm for infinite systems (iDMRG)~\cite{White1992, White1993, McCulloch2008} to variationally optimize the ground state of the model in Eq.~\eqref{equ:hamiltonian} using $U(1)$ charge conservation~\cite{SinghPfeifferVidal2010,SinghPfeifferVidal2011, Hauschild2018}.
To this end, the two-dimensional lattice is wrapped onto a cylinder by closing the boundary condition along its circumference periodically~\cite{Stoudenmire2012}.
We apply the $\mathrm{YC}6$ cylinder geometry~\cite{Hu2019}.
The perturbed ground-state MPS is then time evolved numerically using the matrix-product-operator WII representation of the time evolution operator~\cite{Zaletel2015, Paeckel2019}. To avoid boundary effects, we consider cylinder lengths of $L_x = 51$ for the $120^\circ$ order and the $U(1)$ QSL, $L_x = 81$ in the presence of the chiral interaction, and $L_x = 126$ in the stripe phase. The MPS bond dimension is $\chi_{\mathrm{max}}=2000$. Further details on the numerical procedure are outlined in the End Matter.

Our full dynamical structure factors are complemented by results obtained using the MPS quasiparticle ansatz. This variational approach extends the single-mode approximation and provides direct access to the momentum-resolved excitation spectrum~\cite{haegeman_variational_2012, Vanderstraeten2019, VanDamme2021}. Using $\mathrm{SU}(2)$ symmetry, we can directly target the singlet sector \cite{ZaunerStauberVanderstraeten2018}, so that the dispersion of the lowest-lying singlet excited states can be readily resolved.

\begin{figure*}
    \centering
    \includegraphics[scale=1]{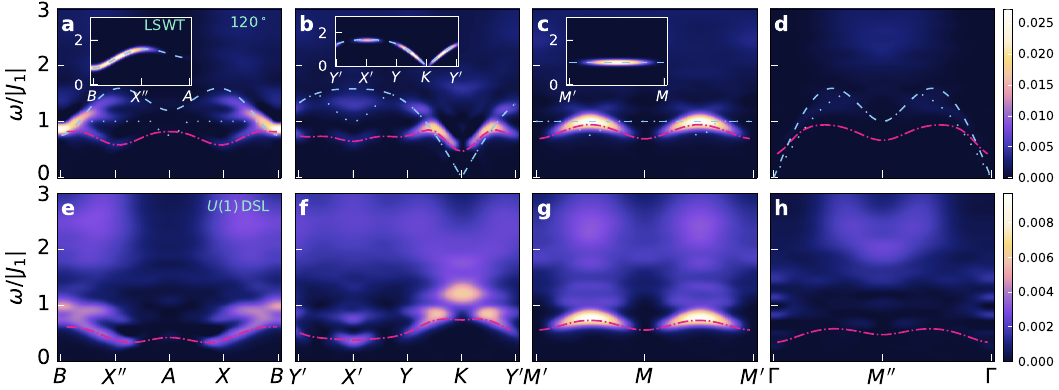}
    \caption{Dynamical dimer structure factor $S_D(\kb, \omega)$ for the $120^\circ$ order (\panelk{a}--\panelk{d}) and the $J_1-J_2$ QSL (\panelk{e}--\panelk{h}).
    The chosen coupling parameters are $J_2 = 0$ and $J_2/J_1 = 0.125$ (both $J_\chi = 0$), respectively.
    The main panels show the results obtained from numerical time evolutions performed on $\mathrm{YC}6$ cylinders with an MPS bond dimension of $\chi = 2000$ and a cylinder length of $L_x = 51$.
    The simulations reached the maximum evolution time of $t_{\mathrm{sim}}=60\,J_1$ in steps of $\delta t = 0.04\,J_1$. The structure factors displayed were obtained for a Gaussian broadening $\sigma = 11.4\,J_1$. The insets in panels \panelk{a}--\panelk{c} for the $120^\circ$ order feature the dimer spectral function along the corresponding cuts as computed within leading-order spin-wave theory. The blue dashed line denotes the magnon dispersion from LSWT. The blue dotted line marks the lower onset of the two-magnon continuum within LSWT. The red dot-dashed line indicates in all panels the lowest-lying excitation obtained from the MPS quasiparticle ansatz.
    }
    \label{fig:main_panel}
\end{figure*}

\paragraph{Ordered phases}%
Figure~\ref{fig:dimer_ssf_all_phases} presents the static dimer structure factor $\chi_D(\k)$ for the ordered phases---$120^\circ$, tetrahedral and stripe order in the panels \panelk{a}, \panelk{c} and \panelk{e}, respectively---and contrasts the numerical MPS results with semiclassical findings from spin-wave theory (\panelk{b}, \panelk{d} and \panelk{f} in the bottom row; see the Supplemental Material~\cite{supp} for details of the calculation).
The main characteristics are reproduced by leading-order spin-wave theory. In the case of the $120^\circ$ phase, the vanishing spectral weight at the $K$ and $M$ points is contrasted with corresponding minima at low amplitude in the static structure factor from MPS. The numerical result exhibits a more detailed structure, e.g., the cut $\Gamma$--$M^{\prime\prime}$ vanishes completely in the spin-wave calculation.
A similar picture holds for the tetrahedral phase. The spectral signal at the $M$ points is minimal both in the numerics and the semiclassical approximation. The $K$ points come with a local minimum of finite intensity in the former result. In both cases, we observe maxima in the proximity of the $X$ points. The diminished weight at the origin in spin-wave theory is a consequence of the perturbative approximation up to leading order.
The analytical expansion gives a finite contribution for the collinear stripe order solely at order $S^2$ for the connected dimer correlations---in contrast to the other two ordered phases. The agreement we find between the numerical static structure factor and the equivalent result within spin-wave theory is conclusive (cf. Fig.~\ref{fig:dimer_ssf_all_phases}\panel{e} and \panel{f}). The momentum $M'$ where the Goldstone mode resides comes with $\chi_D(\k)\big|_{\k\to M'}$ approaching zero.

Figure~\ref{fig:main_panel}\panel{a}--\panel{d} presents the dynamical dimer structure factor $S_D(\k,\omega)$ in the $120^\circ$ phase.
The main panels show the data from the MPS-based time evolution, contrasted with the dynamical spin-wave result in the corresponding insets.
The red dot-dashed line denotes the dispersion obtain from the MPS quasiparticle ansatz. The blue dashed line marks the LSWT magnon branch as a guide~\cite{Chernyshev2009, Drescher2023}.
Along the cut $M$--$M'$, the LSWT result that tracks the single-magnon dispersion is reflected in the low-energy mode of the numerical spectrum, up to a downward renormalization around the midpoints of the Brillouin-zone edges. This closely mirrors the renormalization of the single-magnon mode in the spin DSF~\cite{Zheng2006, Zheng2006PRL, Starykh2006, Chernyshev2006, Oitmaa2020, Sherman2023, Drescher2023}.
Both momentum cuts comprising the various $X$ points exhibit---besides the continua at higher energies---a mode largely following the LSWT result. A significant finite-size gap opens at the corner of the Brillouin zone for the MPS simulation on the cylinder~\cite{Sherman2023, Drescher2023, Drescher2025}.
The distribution of spectral weight is faithfully reflected by the LSWT calculation.
However, whereas the semiclassical approach suggests a dispersion maximum at the $X$ points, both the dynamical data and the quasiparticle ansatz reveal finite spectral weight beneath the onset of the two-magnon continuum. The lowest-lying mode forms a minimum at $X$, $X''$, and $X'$.
This feature can be directly attributed to the phenomenon of avoided quasiparticle decay for the triangular-lattice Heisenberg antiferromagnet~\cite{Ito2017, Verresen2019, MacDougal2020, Drescher2023, Drescher2025}: Due to strong magnon interactions, the magnon branch is partially repelled upon entering the continuum where decay channels are kinematically available.
The observed features of the dimer spectral function arise from the same processes, i.e., the stabilized magnon bands contribute with finite spectral weight to the dynamical dimer signal.
A similar behavior is found in the case of the tetrahedral phase (cf. the spin DSF results in Ref.~\cite{Drescher2025} and Fig.~\ref{fig:tetrahedral_dsf} in the Supplemental Material~\cite{supp}), whereas the spectrum of the stripe order follows qualitatively the predictions from LSWT (Fig.~\ref{fig:stripe_dsf}).

\paragraph{Quantum spin-liquid regimes}%
The ground states found via iDMRG routines for the candidate spin-liquid regimes---the quantum disordered domain for $0.07\lesssim J_2/J_1 \lesssim0.15$ on the $J_2$ axis and the CSL for finite $J_\chi$---do not exhibit a dominant spin order (cf. explicitly Ref.~\cite{Drescher2025}). For the $J_1-J_2$ QSL, we consider the distinct lower-energy sector on the $\mathrm{YC}6$ cylinder~\cite{Zhu2015, Hu2015, Hu2019, Drescher2023, Drescher2025, JiangWhite2026, KovalskaDelft2026}, while we stay in the trivial sector of the CSL with a non-degenerate entanglement spectrum exposing the phase's chiral nature~\cite{Wen1991, LiHaldane2008, Szasz2020, Cookmeyer2021, Kadow2022, Kuhlenkamp2024, Drescher2025}.
Figure~\ref{fig:dimer_ssf_all_phases}\panel{g} and \panel{h} show the static dimer structure factor related to the two phases, respectively.
The peak amplitudes are substantially smaller than in the ordered phases.
While both the $J_1-J_2$ QSL and the CSL share common features---such as the reduced spectral intensity at the $M$ points, reflecting the influence of adjacent ordered phases---the behavior at the $K$ points highlights clear distinctions from the ordered regimes. In the former spin liquid, the spectral weight at $K$ increases significantly compared with the $120^\circ$ order, while in the CSL a distinct maximum develops at the same momentum.

The dimer DSF for the candidate $J_1-J_2$ QSL presented in the bottom row of Fig.~\ref{fig:main_panel} reveals similarities along the cut $M$--$M'$ with the $120^\circ$ phase. The excitation gap at the $M$ points though is reduced, as in the spin DSF~\cite{Sherman2023, Drescher2023, Drescher2025}. For the field theory of a DSL, the excitations at these momenta are expected to form gapless continua in the thermodynamic limit~\cite{Song2019, Song2020, willsher_dynamics_2025}. The lowest mode as probed by the dimer operator in Fig.~\ref{fig:main_panel}\panel{g} exposes vanishing spectral intensity, with remnants of spectral weight being present in the continuum at higher energies. The continuum is much more pronounced in the liquid regime than in the coplanar order. It reflects the presence of fractional spinon excitations whose combinations contribute both to the singlet and the triplet channels.

Unlike the centers of the Brillouin-zone edges, the features at $K$ differ markedly between the two adjacent phases. In the $120^\circ$ ordered domain, the low-energy signal forms a minimum at this momentum with regard to both the dispersion and the spectral weight.
In the QSL, by contrast, we observe a pronounced horizontal feature at intermediate energies, while the lower edge of the excitation spectrum shifts upward as $K$ is approached. Both characteristics are readily captured by the MPS quasiparticle ansatz.
Thus, the spectrum at $K$ evidently distinguishes the $120^\circ$ order from the candidate QSL, a distinction that is not apparent with comparable clarity in the spin correlations~\cite{Sherman2023, Drescher2023, Drescher2025, JiangWhite2026, KovalskaDelft2026}.
Moreover, the spinon dispersion for a gapped $\mathbb{Z}_2$ QSL, which comes with minima at the $K$ points~\cite{Sachdev1992, Wang2006}, would necessarily result in a singlet response comprising two-spinon states that has an absolute excitation minimum at $K$ as well. This contradicts our results. The softening of the gap at the $M$ points, however, is compatible with the spinon structure of the $U(1)$ DSL~\cite{willsher_dynamics_2025}.

Most importantly, however, we find a distinct minimum in the low-lying excitation at $X'$ along with---in a slightly shifted manifestation caused by the anisotropies of the cylinder simulation (cf. Fig.~\ref{fig:dimer_ssf_all_phases}\panel{g})---the momenta $X$ and $X''$. Contrary to the dispersion minima at these momenta in, e.g., the $120^\circ$ ordered and tetrahedral phases (Figs.~\ref{fig:main_panel}\panel{b} and \ref{fig:tetrahedral_dsf}\panel{b}) or the CSL regime (Fig.~\ref{fig:csl_panel}\panel{b}), the excitation gap at $X'$ in the putative spin liquid constitutes the lowest-lying excitation with finite weight across all accessible momenta.
Compared to the $120^\circ$ phase, the spectral weigth at the minimum at $X'$ is enhanced relative to other momenta.
This is naturally accounted for by the $\mathrm{QED}_3$ field theory scenario:
while the triplet monopole excitations of the DSL on the triangular lattice reside at the corners of the Brillouin zone, the singlet monopoles---invisible in the spin correlations but detectable in the dimer response---are located precisely at the $X$ points~\cite{Song2019, Song2020}.
In the two-dimensional limit, these excitations are expected to manifest as conical gapless continua~\cite{willsher_dynamics_2025}.
These low-energy continua cannot be resolved numerically on the accessible system sizes.
Instead, a gapped low-energy excitation emerges at $X$. We note that this finite-size gap is analogous to the behavior seen in the spin structure factor at the Brillouin-zone boundaries, where both expected gapless continua at the $K$ and $M$ points have similar gap sizes and are further dominated by fluctuations related to the nearby ordered states (cf. Ref.~\cite{willsher_dynamics_2025}).
The observed finite-size gap is comparable to the corresponding result obtained from variational Monte Carlo simulations~\cite{Budaraju2025}.
The resulting spectral features---particularly at the decisive $X$ points---provide support for interpreting the phase as an ultimately gapless DSL~\cite{Song2019, Song2020, willsher_dynamics_2025}.

\paragraph{Outlook}%
Leveraging GPU-accelerated large-scale MPS simulations, we obtain high-resolution spectra of the dynamical dimer response, paving a new route to probe excitations above exotic ground states in frustrated magnets. The approach yields distinctive signatures that enable discrimination between competing field theories and connects directly to experiments, including momentum-resolved RIXS and Raman scattering, thereby providing a valuable framework for quantitative studies of antiferromagnetic models with QSL regimes beyond analytic control. The resulting spectra serve as a direct guide for future experiments, indicating which features to target and how to distinguish between underlying theoretical descriptions.

\paragraph{Data availability}%
All data are available in the Zenodo repository~\cite{zenodo2026}.

\paragraph{Acknowledgments}%
We thank Sylvain Capponi, Eduard Koller, Andreas Läuchli, Alexander Wietek and Josef Willsher for helpful discussions.
M.D. acknowledges Jonas Habel for 
valuable input on the spin-wave calculations.  
J.K. and F.P. acknowledge support from TRR 360 - 492547816, the Deutsche Forschungsgemeinschaft (DFG, German Research Foundation) under Germany’s Excellence Strategy EXC-2111-390814868, as well as the Munich Quantum Valley, which is supported by the Bavarian state government with funds from the Hightech Agenda Bayern Plus.
J.K. also acknowledges support from DFG grants No. KN1254/1-2, KN1254/2-1.
R.M. recognizes support by
the Deutsche Forschungsgemeinschaft under Grant No. SFB
1143 (Project-ID No. 247310070) and by
the Deutsche Forschungsgemeinschaft  under cluster of excellence
ctd.qmat (EXC 2147, Project-ID No. 390858490).

\bibliography{references_dimer_dsf}

\clearpage

\appendix
\onecolumngrid
    \section*{End Matter}
\twocolumngrid

\appsection{Details on the numerical procedure and the ground state optimization}
\label{sec:numerical_details}

For the ground state optimization, we apply the density matrix renormalization group algorithm with infinite boundary conditions (iDMRG). Concretely, we use the $\mathrm{YC}6$ cylinder geometry, indicating that the system is invariant under translation by $L\cdot \mathbf{a}_2 = 6 \,\mathbf{a}_2$ with the Bravais vectors $\mathrm{a}_2 = (1/2, \sqrt{3}/2)^T$ and $\mathrm{a}_1=(1,0)^T$ (cf. Fig.~\ref{fig:schematic_phase_diagram}\panel{b}). The iDMRG variational optimization is performed on a unit cell with three rings of circumference $L=6$ and a maximum bond dimension of $\chi = 2000$.
We apply Abelian charge conservation~\cite{Hauschild2018, Hauschild2024}, i.e., the total $S_z$ quantum number is preserved during the simulation.
Whereas the ground state representations of the ordered phases are stabilized straight-forwardly on the cylinder, the paramagnetic regimes exhibit two distinct topological sectors on cylinders with an even number of legs.
In the case of the $J_1-J_2$ quantum spin liquid (QSL), we focus on the energetically favorable odd sector~\cite{Zhu2015, Hu2015, Hu2019}---or low-energy sector~\cite{JiangWhite2026}---that exhibits a more isotropic correlation structure.
In the chiral spin-liquid phase, we perform the dynamics in the directly accessible trivial sector that comes with a non-degenerate entanglement spectrum consistent with the Wess-Zumino-Witten conformal field theory description of the gapless edge mode (cf. Ref.~\cite{Szasz2020, Cookmeyer2021}).
The ground state approximation used for the quasiparticle excitation ansatz~\cite{haegeman_variational_2012, Vanderstraeten2019} was obtained from the variational uniform MPS (VUMPS) algorithm~\cite{HaegemanVerstraete2017, ZaunerStauber2018}, featuring $\mathrm{SU}(2)$-symmetric tensors.

The MPS with infinite boundary conditions that represents the ground state of the system for given set of parameters can be used to build enlarged unit cells by periodically stacking them. These systems serve as the basis for the dynamical evolution. For improved computational efficiency in cutting-edge simulations, it is advantageous to terminate the infinite boundary conditions by a suitable eigenvector, thereby mapping them to a finite system with open boundary conditions (cf. Refs.~\cite{Drescher2023, Drescher2025}). 

The time evolution is initialized by applying a single-bond operator
\begin{equation}
    \hat{\mathbf{S}}_i \cdot \hat{\mathbf{S}}_{i + \delta} - \braket{\hat{\mathbf{S}}_i \cdot \hat{\mathbf{S}}_{i + \delta}}
    \label{equ:bond_operator}
\end{equation}
onto the nearest-neighbor bond $(\vr_i,\, \vr_i + \bm{\delta})$ with $\bm{\delta} \in \{\pm \mathbf{a}_1, \pm \mathbf{a}_2, \pm \mathbf{a}_3 = \pm (\mathbf{a}_1- \mathbf{a}_2)\}$ denoting the displacement vector of the nearest-neighbor pairs.
We choose a site $j_c$ in the center of the system and prepare distinct initial wave functions by applying the bond operator on bond $(j_c, j_c + \delta)$ for all choices of $\delta$. In order to prevent numerical artifacts arising from the anisotropies of the cylinder setup, we do not impose assumptions about the rotation symmetries of the system, but perform all time evolutions independently. As the time-step size, we use $\delta t = 0.04\,J_1$. After each $N_{\mathrm{steps}}=5$ time steps, we compute the time-dependent dimer-dimer correlation.
To avoid numerical artifacts---such as energy shifts causing spurious phases in the wave function due to Trotter errors and finite bond-dimension effects in the variational application of the MPO to the MPS---we simultaneously time-evolve the unperturbed ground state and use this reference state to compute the correlations. Eventually, the bond correlations from the six individual time evolutions (and the additional ground state evolution) can be combined to obtain the full dynamical dimer-dimer correlations for the operator in Eq.~\ref{equ:dimer_operator}.
Note that the subtraction of the expectation value in Eq.~\eqref{equ:dimer_operator} and the application of the full bond operator in Eq.~\ref{equ:bond_operator} onto the ground state in the form of an MPO proved essential for the numerical accuracy of the simulations.

Having obtained the dynamical dimer correlation function, we can invoke the standard recipe to transform the data into the corresponding dynamical structure factor. After Fourier transforming the spatial indices to momentum space (cf. Fig.~\ref{fig:schematic_phase_diagram}\panel{c} for the accessible momentum cuts on this cylinder geometry), we apply a linear extrapolation of the time series for each momentum and find eventually the spectral function $S_D(\k,\omega)$ from a convolution of the signal with a Gaussian window function.
We choose a value $\alpha$ of the Gaussian envelope at the maximum evolution time $t_{\mathrm{sim}}$, defining the broadening via $\sigma = \sqrt{t_{\mathrm{sim}}^2/(-2\ln(\alpha))}$. Typical values are for instance $\alpha = 10^{-6}$, i.e., $\sigma \approx 11.4\,J_1$ for $t_{\mathrm{sim}} = 60\,J_1$.
In order to improve the resolution in momentum space, we can apply an interpolation scheme prior to performing the temporal Fourier transformation (for details consult Ref.~\cite{Drescher2025}), which has been applied for the numerical data presented in this work.

The large-scale dynamical simulations have been performed on a cluster setup that combines four NVIDIA A100 GPUs (80 GB memory each) and 1 TB of system memory on each node alongside two Intel Xeon Platinum 8360Y CPUs (36 cores at 2.40GHz).
We use a self-developed tensor-network library that stores the data tensors in sparse form by exploiting the $U(1)$ symmetry. Whenever possible, the tensor blocks are retained in GPU memory, subject to the available memory capacity.
We apply a memory management scheme that transfers data asynchronously between the system memory and the GPUs if necessary. In the $120^\circ$ phase and the candidate QSL at $J_2/J_1 = 0.125$, we coupled two GPUs per simulation. In the other cases, we used four GPUs in parallel and kept data such as site tensors and contracted environments that were not immediately required in the main memory.

\clearpage
\makeatletter
\def\ps@mypagestyle{%
  \def\@oddhead{}%
  \def\@evenhead{}%
  \def\@oddfoot{\hfil\thepage}%
  \def\@evenfoot{\hfil\thepage}%
}
\makeatother

\pagestyle{mypagestyle}
\setcounter{page}{1}

\onecolumngrid
    \section*{Supplemental Material}
\twocolumngrid

\begin{figure}
    \centering
    \includegraphics[scale=1]{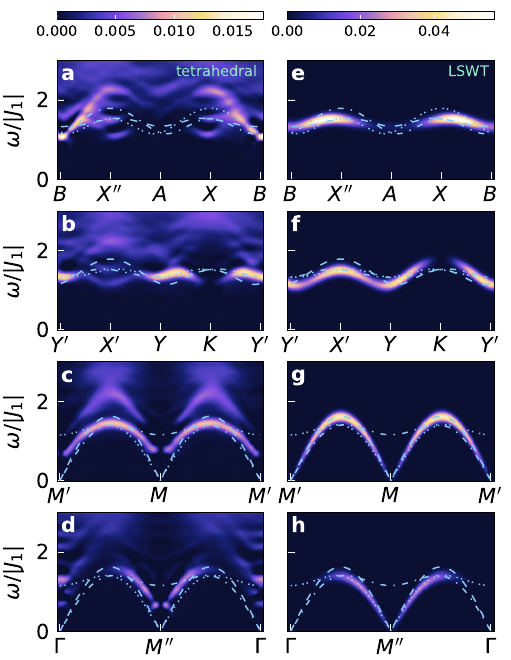}
    \caption{The dynamical dimer structure factor in the tetrahedrally ordered phase at $J_\chi/J_1 = 0.5$ and $J_2/J_1 = 0.125$.
    The left column shows the results from dynamical MPS simulations at bond dimension $\chi = 2000$ on a $\mathrm{YC}6$ cylinder of size $6\times 81$. The maximum evolution time was $t_{\mathrm{sim}} = 60\,J_1$ with a time-step size of $\delta t = 0.04\,J_1$. The right-hand column presents the leading-order dynamical spin-wave results obtained on a cylinder of dimensions $6 \times 220$.
    The spectral function was computed with a Gaussian broadening determined by $\alpha = 10^{-6}$.
    The blue lines indicate the three magnon modes arising from the various ground state configurations (cf. the section on spin-wave theory).
    }
    \label{fig:tetrahedral_dsf}
\end{figure}

\begin{figure}
    \centering
    \includegraphics[scale=1]{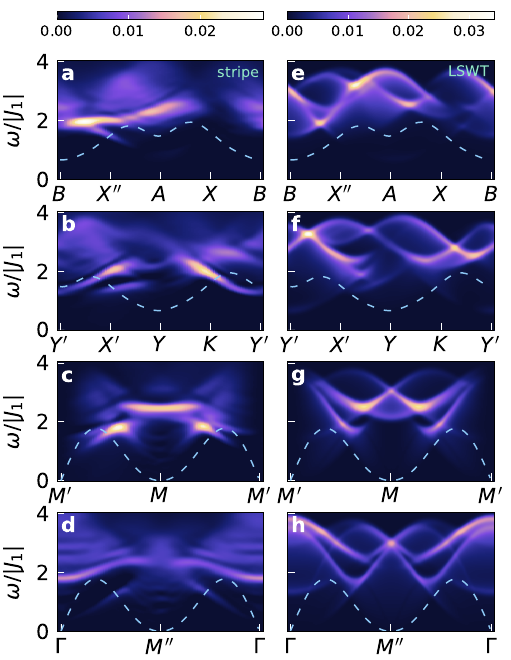}
    \caption{The dynamical dimer structure factor in the stripe phase at $J_2/J_1 = 0.55$ ($J_\chi = 0$).
    The spectral function cuts from the large-scale MPS numerics is displayed on the left. The simulation was performed at a bond dimension of $\chi = 2000$ on a $\mathrm{YC}6$ cylinder with $L_x = 126$. The dynamical parameters are identical to in Fig.~\ref{fig:tetrahedral_dsf}.
    The right column shows the corresponding result from the lowest non-vanishing order in spin-wave theory. It was computed on an equivalent cylinder of size $6\times 180$.
    The blue dashed line marks the single-magnon dispersion as obtained within LSWT.
    }
    \label{fig:stripe_dsf}
\end{figure}

\begin{figure*}
    \centering
    \includegraphics[scale=1]{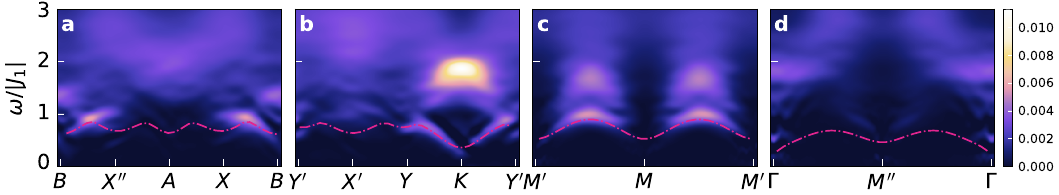}
    \caption{The dynamical dimer structure factor in the chiral spin-liquid phase.
    The chosen coupling parameters are $J_2/J_1 = 0.125$ and $J_\chi/J_1 = 0.2$.
    The main panels show the results obtained from numerical time evolutions performed on $\mathrm{YC}6$ cylinders with an MPS bond dimension of $\chi = 2000$ and a cylinder length of $L_x = 81$.
    The simulations were performed up to $t_{\mathrm{sim}}=60\,J_1$ in steps of $\delta t = 0.04\,J_1$. We applied a Gaussian broadening of $\sigma = 11.4\,J_1$. The red dot-dashed line marks the excitation dispersion obtained from the MPS quasiparticle ansatz.
    }
    \label{fig:csl_panel}
\end{figure*}

\appsection{Dynamic dimer structure factor in additional phases}
Figure~\ref{fig:csl_panel} summarizes the findings for the dynamical dimer structure factor in the chiral spin-liquid phase at $J_\chi/J_1 = 0.2$ ($J_2/J_1 = 0.125$).
The chiral spin liquid is an inherently gapped phase~\cite{Wen1989, Wen1991}.
Accordingly, the lower edge of the excitation continuum along $M$--$M'$ rises with the onset of the chiral coupling $J_\chi$ (Fig.~\ref{fig:main_panel}\panel{g} vs. Fig.~\ref{fig:csl_panel}\panel{i}).
The continuum strength across all momentum cuts resembles that of the candidate DSL phase.
The most prominent feature appears at relatively high energies around $K$: it manifests as a horizontal structure of strong spectral intensity within the continuum and can be associated with a two-particle combination of the spin-$1$ bound state at the same momentum identified in Ref.~\cite{Drescher2025}, consistent with the predictions for the Kalmeyer-Laughlin collective excitation~\cite{Kalmeyer1989}.
Previous work on the spin structure factor reported spurious low-energy spectral weight at the high-symmetry momenta $X$, $M$, and $K$~\cite{Drescher2025}, consistent with the emergence of two-spinon continua within the Kalmeyer-Laughlin ansatz.
In accordance with this expectation, both the dynamical simulation and the quasiparticle ansatz reveal faint low-lying spectral weight at the $M$ points and a clear mode---possibly a singlet two-spinon bound state---with its minimum precisely at $K$.
The spectral function in the chiral spin-liquid phase displays overall more anisotropies and truncation-related artifacts as the corresponding MPO bond dimension is significantly larger than in the absence of the chiral interaction term.

Figures~\ref{fig:tetrahedral_dsf} and \ref{fig:stripe_dsf} finally present the dynamical dimer structure factor in the tetrahedral and the stripe ordered phases, respectively.
We compare the numerical result (in the left column) with the spin-wave calculation to leading order on the right.
As described in the main text, the main spectral features in the tetrahedral phase are captured by the LSWT result. Modifications are caused by finite-size gaps at the $M$ points and downward renormalizations of the contributions from magnon branches in the vicinity of the corner points $K$.
The continua arising from the MPS simulations cannot be captured by the lowest-order spin-wave computation.
Most importantly, the $X$ points exhibit low-energy modes (cf. for instance Fig.~\ref{fig:tetrahedral_dsf}\panel{b}) that emerge below the onset of the two-magnon continuum in LSWT. The lower edge of this continuum is marked by the minimum of the three distinct magnon modes for each momentum that are indicated by the blue lines of various patterns.
Like in the $120^\circ$ order, we can connect this low-lying spectral weight to the repulsion of magnons from the continuum due to strong interaction processes---a phenomenon known as avoided quasiparticle decay~\cite{Verresen2019}. 

The structure of the spectra in the stripe order, on the other hand, are reproduced in good agreement by the semiclassical approximation. Note that for the collinear order, contribution to the dimer structure factor of order $S^3$ vanishes completely. The next order $S^2$, acting as the leading-order signal in this domain, includes two-particle continua. The overall distribution of spectral weight and the dominant features within the continua demonstrate convincing agreement between the two approaches. As a guide to the eye, we also plot the single-magnon dispersion from LSWT on top of the full spectral function.

\appsection{Profile plots of the static dimer structure factor}

\begin{figure}
    \centering
    \includegraphics[scale=1]{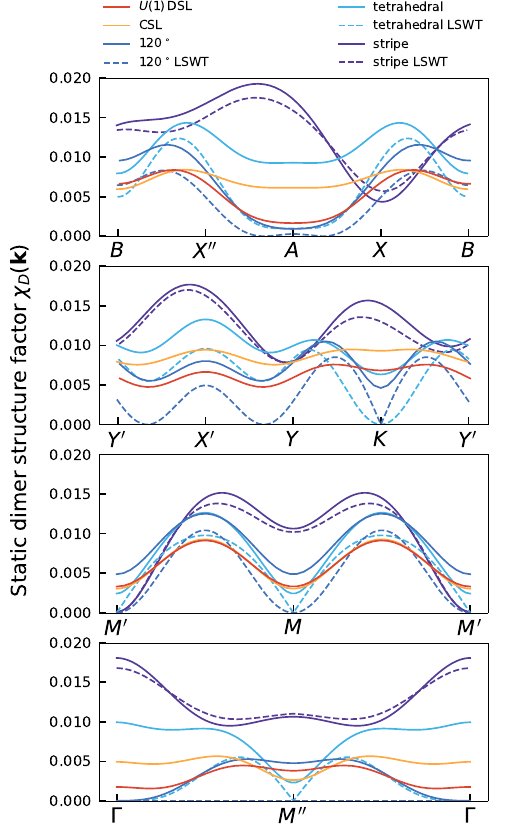}
    \caption{Profile plots of the static dimer structure factor $\chi_D(\kb)$ for the various phases considered in this work. For the ordered ground states, we present also the leading-order spin-wave result as a dashed line of the corresponding color. The data displayed and hence also the parameter choices are identical to Fig.~\ref{fig:dimer_ssf_all_phases} in the main text.}
    \label{fig:ssf_profiles}
\end{figure}

Figure~\ref{fig:ssf_profiles} presents the data for the static dimer structure factor
\begin{equation}
    \chi_D(\k) = \frac{1}{N} \sum_{i,j} e^{-\ui \k \cdot(\r_i - \r_j)} \braket{\hat{\mathcal{D}}_i \hat{\mathcal{D}}_j}\,,
\end{equation}
for the dimer operator $\hat{\mathcal{D}}_i$ as defined in the main text for the various ordered and liquid phases: the candidate $U(1)$ Dirac spin liquid (DSL) and the chiral spin liquid (CSL), the $120^\circ$, tetrahedral and stripe orders. In contrast to Fig.~\ref{fig:dimer_ssf_all_phases} in the main text, we plot here the intensity profiles along the distinct cuts that are accessible on the $\mathrm{YC}6$ cylinder geometry.

In the ordered phases, we compare the numerical MPS result (solid lines) to the leading-order spin-wave contribution (dashed line of the same color). The results for the stripe order at $J_2/J_1 = 0.55$ agree very closely between the numerical approach and the analytical approximation. For the $120^\circ$ order and the tetrahedral phase, we observe a qualitative agreement of the main features. In general, spectral weight that is captured in the numerical finding is not present in the semiclassical result due to the missing higher-order contributions. The intensity along the cut $M''$--$\Gamma$ for instance vanishes completely within linear spin-wave theory (LSWT) in the coplanar $120^\circ$ phase.
The disordered regimes generally exhibit a reduced maximum amplitude of $\chi_D(\kb)$ compared to the ordered domains.

As an estimate for the anisotropies caused by the cylinder geometry, we can compare the amplitude at the various edge centers of the Brillouin zone. The Goldstone mode at $M'$ in the stripe phase is reflected by vanishing weight at this momentum only. The $120^\circ$ order would exhibit a sixfold rotation symmetry in the thermodynamic limit. The signals at $M$ and $M'$ versus $M''$ agree very well (solid dark blue line). The same holds for the tetrahedral order, but with a significantly reduced value at the $M$ points (solid light blue line), related to the presence of Goldstone modes at these momenta in this phase.
The candidate quantum spin-liquid regimes come with more pronounced differences: Whereas at $M'$ and $M$, the curves for the DSL and CSL are almost identical, they signals split up at $M''$. The CSL acquires a slightly lower intensity at $M''$ compared to the other two edge centers, the DSL signal on the other hand undergoes a small upward shift.

We also observe that the Goldstone excitations at $K$ in the $120^\circ$ phase and at the $M$ points in the tetrahedral phase, characterized by a linear dispersion when approaching the corresponding momenta, exhibit a finite-size gap in the numerical MPS-based result.

\appsection{Spin-wave theory}
\label{sec:lswt_supp}

\appsubsection{General solution of the energy dispersion in linear spin-wave theory}

The semiclassical spin-wave approximation assumes that in an magnetically ordered ground state of a spin system, quantum fluctuations cause also slight deviations from the classical spin configuration. These deviations can be treated perturbatively. In this paragraph, we want to develop a general form of the resulting Hamiltonian and its solution within linear spin-wave theory.
Starting from a spin Hamiltonian such as in Eq.~\ref{equ:hamiltonian}, the canonical Holstein-Primakoff transformation allows us to transition from the spin operators (that are aligned with the z-axis of a locally rotated reference frame) to bosonic operators:
\begin{align}
    \mathcal{S}_i^z &= S - \hat{a}_i^\dagger \hat{a}_i^{\phantom{\dagger}}\\
    \mathcal{S}_i^+ &= \sqrt{2S} \,\sqrt{1-\frac{\hat{a}_i^\dagger\hat{a}_i^{\phantom{\dagger}}}{2S}} \hat{a}_i^{\phantom{\dagger}} \approx  \sqrt{2S} \,\hat{a}_i^{\phantom{\dagger}}\\
    \mathcal{S}_i^- &= \sqrt{2S} \, \hat{a}_i^\dagger\sqrt{1-\frac{\hat{a}_i^\dagger\hat{a}_i^{\phantom{\dagger}}}{2S}} \approx  \sqrt{2S} \,\hat{a}_i^{\dagger}
\end{align}

The local spin operators $\mathcal{S}_i^\alpha$ are aligned with the z-axis of the local frame and are related to the spin operator $\hat{\mathbf{S}}_i$ in the global frame by rotation matrices $R_i$ via
\begin{equation}
    \hat{\mathbf{S}}_i = R_i \mathcal{S}_i\,.
    \label{equ:rotated_frames}
\end{equation}
Here, $\locS_i = (\locS_i^x, \locS_i^y, \locS_i^z)^T$ is the vector of spin operators in the local frame at site $i$. 
Generally, we find a transformed Hamiltonian in terms of the bosonic operators that acquires a form as follows:
\begin{equation}
    \hat{H} = H_0 + \sum_i A_i \hat{a}_i^\dagger \hat{a}_i^{\phantom{\dagger}}
    + \sum_{(i,j)} \left[C_{ij} \hat{a}_i^\dagger \hat{a}_j^{\phantom{\dagger}} + D_{ij} \hat{a}_i^{\phantom{\dagger}} \hat{a}_j^{\phantom{\dagger}} + \mathrm{h.c.}\right]\,.
    \label{eq:H2_real_space}
\end{equation}
Here, $H_0$ and the $A_i$ are real constants and the sum $\sum_{(i,j)}$ runs over pairs of sites $(i,j)$ that can be assumed to be distinct, i.e., $i \ne j$.

We can now Fourier transform the operators from discrete spatial coordinates $\vr = \vr_i \,\hat{=}\, i$
to momentum space as
\begin{equation}
    \ann_\vr = \frac{1}{\sqrt{N}} \sum_\kb e^{-\ui \kb \cdot \vr} \ann_\kb\,.
\end{equation}
The resulting Hamiltonian in quadratic order can be brought to the following form:
\begin{equation}
    H_2 = \sum_{\k} \left[A_{\k}^{\phantom{*}} \hat{a}_\k^\dagger \hat{a}_\k^{\phantom{\dagger}} - \frac{1}{2} \left[B_\k^* \hat{a}_\k^\dagger \hat{a}_{-\k}^{\dagger} + B_\k^{\phantom{*}} \hat{a}_{-\k}^{\phantom{\dagger}} \hat{a}_\k^{\phantom{\dagger}}\right]\right]\,.
    \label{eq:H2_momentum_space}
\end{equation}

The coefficients derive directly from the hopping constants in real space in Eq.~\eqref{eq:H2_real_space} via
\begin{align}
A_\kb &= A + \sum_{\vd} 2 \,\mathfrak{Re}\left\{C_{\vd} e^{-\ui \kb \cdot \vd}\right\}\,,\\
B_\kb &= - \sum_{\vd} 2\,D_{\vd} \, \cos\left(\kb \cdot \vd\right)\,.
\end{align}
We assume here that the coefficients $A_i \equiv A \in \mathbb{R}\forall i$ are site independent and that $C_{\vd}$, $D_{\vd}$ only depend on the distance vector $\vd = \vr_j - \vr_i$ between two sites, not the position of the sites within the unit cell of the translationally invariant lattice, i.e., they need to be independent of the sublattice index of the magnetic order.
In general, the resulting $A_\kb$ are real while the $B_\kb$ can be complex.
One can always symmetrize the momentum-dependent coefficients such that they become even functions of momentum $\kb$: $A_{\kb} = A_{-\kb}$ and $B_\kb = B_{-\kb}$.

The Bogoliubov transformation
\begin{align}
    \hat{a}_\k^{\phantom{\dagger}} &= u_\k^{\phantom{*}} \hat{b}_\k^{\phantom{\dagger}} + v_\k^* \hat{b}_{-\k}^\dagger \\
    \hat{a}_{-\k}^\dagger &= u_\k^* \hat{b}_{-\k}^\dagger + v_\k^{\phantom{*}} \hat{b}_\k^{\phantom{\dagger}}
\end{align}
allows us to diagonalize the quadratic Hamiltonian.

The linear spin-wave dispersion then yields
\begin{equation}
    \varepsilon_\k = \sqrt{A_\k^2 - \left|B_\k\right|^2}\,.
\end{equation}
Writing the coefficient $B_\k$ in polar form as $B_\k = \left|B_\k\right| e^{\ui \varphi_\k}$, the Bogoliubov coefficients are given as
\begin{align}
    u_\k &= \sqrt{\frac{A_\k + \varepsilon_{\k}}{2\varepsilon_\k}}\,,\\
    v_\k &= e^{\ui \varphi_\k} \sqrt{\frac{A_\k - \varepsilon_\k}{2\varepsilon_\k}}\,.
\end{align}

Hence, the quadratic Hamiltonian reduces to 
\begin{equation}
    H_2 = \sideset{}{'}\sum_\k \varepsilon_\k \left(\hat{b}_\k^\dagger \hat{b}_\k^{\phantom{\dagger}} + \frac{1}{2} \right) - \sum_\k \frac{A_\k}{2}\,.
\end{equation}
The primed sum indicates that the Goldstone modes are excluded from the summation. For those momenta, the Bogoliubov transformation is not well defined.

\begin{figure}
    \centering
    \begin{minipage}{\linewidth}
        \centering
        \LatticeTetrahedral
    \end{minipage}

    \vspace{1em}
    
    \begin{minipage}{0.48\linewidth}
        \centering
        \LatticeTetrahedralTwo
    \end{minipage}
    \begin{minipage}{0.48\linewidth}
        \centering
        \LatticeTetrahedralThree
    \end{minipage}
    \caption{Spin configurations for the tetrahedral order. A possible choice for the classical spin orientations on the four sublattices is given in Eqs.~\eqref{eq:6_tetrahedral_spins_1}-\eqref{eq:6_tetrahedral_spins_2}.}
    \label{fig:tetrahedral_arrangements}
\end{figure}

\begin{figure}
    \centering
    \includegraphics[scale=1]{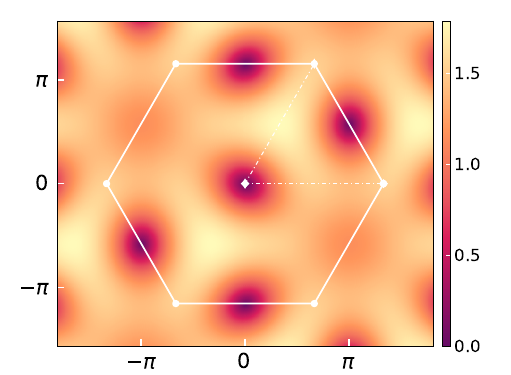}
    \caption{The magnon dispersion $\varepsilon_\kb$ for the tetrahedral order (configuration 1) on the triangular lattice. We consider a spin-\nicefrac{1}{2} system with couplings $J_2/J_1 = 0.125$ and $J_\chi/J_2 = 0.5$. The dispersions for configurations 2 and 3 are rotated clockwise by $\pi/3$ and $2\pi/3$, respectively.}
    \label{fig:dispersion_tetrahedral}
\end{figure}

\paragraph{$\mathit{120}^\circ$ order}%
In the coplanar $120^\circ$ order, it suffices to apply rotations around the $y$-axis locally:
\begin{align}
S^x_i & = \sin(\vartheta_i) \locS_i^z + \cos(\vartheta_i) \locS_i^x\,, 
\label{eq:6_local_rotation_Sx}\\
S^z_i &= \cos(\vartheta_i) \locS_i^z - \sin(\vartheta_i) \locS_i^x\,.
\label{eq:6_local_rotation_Sz}
\end{align}
The classical ground state energy is
\begin{equation}
    H_0 = - \frac{3}{2}J_1 S^2 N\,.
\end{equation}
The momentum-dependent coefficients of the Hamiltonian read
\begin{align}
    A_\kb &= 3 J_1 S \left(1 + \frac{1}{2} \gamma_\kb \right)\,, \\
    B_\kb &= \frac{9}{2} J_1 S \gamma_\kb\,,
\end{align}
where we introduced the quantity~\cite{Chernyshev2009}
\begin{align}
    \gamma_\kb &= \frac{1}{6} \sum_{\vd} \left(e^{i \kb \cdot \vd} + e^{-i \kb \cdot \vd}\right) \\
    &= \frac{1}{3} \left[\cos(k_x) + 2  \cos\left(\frac{k_x}{2}\right) \cos\left(\frac{\sqrt{3} k_y}{2}\right) \right]\,.
\end{align}
The $\vd$ iterate over all nearest-neighbor displacements.
The magnon dispersion consequently yields
\begin{equation}
    \varepsilon_\kb = \sqrt{A_\kb^2 - B_\kb^2} = 3 J_1 S \sqrt{\left(1 - \gamma_\kb\right) \left(1 + 2 \gamma_\kb\right)}\,.
\end{equation}

\paragraph{Stripe order}%
The collinear alternating stripe order gives a classical ground state energy of
\begin{equation}
    H_0 = - S^2 N (J_1 + J_2)\,.
\end{equation}
In this case, we consider both nearest- and next-nearest-neighbor interactions ($J_2/J_1 \gtrsim 0.15$).
The coefficients in Eq.~\eqref{eq:H2_momentum_space} are given as
\begin{align}
    A_\kb &= 2J_1 S \left[1 + \cos(\kb \cdot \mathbf{a}_1) \right] + 2 J_2 S \left[1 + \cos(\kb \cdot \mathbf{l}_1)\right] \\
    B_\kb &= 2J_1 S \left[\cos(\kb \cdot \mathbf{a}_2) + \cos(\kb \cdot \mathbf{a}_3)\right] \nonumber \\
    &+ 2 J_2 S \left[\cos(\kb \cdot \mathbf{l}_2) + \cos(\kb \cdot \mathbf{l}_3)
    \right]\,.
\end{align}
Besides the nearest-neighbor vectors $\vd$, the next-nearest-neighbor vectors denoted as $\mathbf{l}_1 = 2\mathbf{a}_2 - \mathbf{a}_1$, $\mathbf{l}_2 = \mathbf{a}_1 + \mathbf{a}_2$, and $\mathbf{l}_3 = 2\mathbf{a}_2 - \mathbf{a}_1$ occur in the expressions.

\begin{widetext}
\paragraph{Tetrahedral order}%
In the tetrahedral order, we have four sublattices with the classical spin orientations of the four sites in a unit cell pointing towards the corners of a tetrahedron. A suitable choice for the classical spin orientations is
\begin{align}
    \mathbf{S}_A &= \frac{S}{\sqrt{3}} \begin{pmatrix} 1 \\ 1 \\ 1 \end{pmatrix}\,, \quad
    \mathbf{S}_B = \frac{S}{\sqrt{3}} \begin{pmatrix} 1 \\ -1 \\ -1 \end{pmatrix}\,, \label{eq:6_tetrahedral_spins_1}\\
    \mathbf{S}_C &= \frac{S}{\sqrt{3}} \begin{pmatrix} -1 \\ 1 \\ -1 \end{pmatrix}\,, \quad
    \mathbf{S}_D = \frac{S}{\sqrt{3}} \begin{pmatrix} -1 \\ -1 \\ 1 \end{pmatrix}\,
    \label{eq:6_tetrahedral_spins_2}
\end{align}
with the corresponding rotation matrices according to Eq.~\ref{equ:rotated_frames} given as
\begin{align}
    R_A &= \begin{pmatrix}
        \frac{1}{\sqrt{2}} & \frac{1}{\sqrt{6}} & \frac{1}{\sqrt{3}} \\
        -\frac{1}{\sqrt{2}} & \frac{1}{\sqrt{6}} & \frac{1}{\sqrt{3}} \\
        0 & -\frac{2}{\sqrt{6}} & \frac{1}{\sqrt{3}}
    \end{pmatrix} &
    R_B &= \begin{pmatrix}
        \frac{1}{\sqrt{2}} & \frac{1}{\sqrt{6}} & \frac{1}{\sqrt{3}} \\
        \frac{1}{\sqrt{2}} & -\frac{1}{\sqrt{6}} & -\frac{1}{\sqrt{3}} \\
        0 & \frac{2}{\sqrt{6}} & -\frac{1}{\sqrt{3}}
    \end{pmatrix} \label{eq:6_tetrahedral_frames_1}\\
    R_C &= \begin{pmatrix}
        \frac{1}{\sqrt{2}} & \frac{1}{\sqrt{6}} & -\frac{1}{\sqrt{3}} \\
        \frac{1}{\sqrt{2}} & -\frac{1}{\sqrt{6}} & \frac{1}{\sqrt{3}} \\
        0 & -\frac{2}{\sqrt{6}} & -\frac{1}{\sqrt{3}}
    \end{pmatrix} &
    R_D &= \begin{pmatrix}
        \frac{1}{\sqrt{2}} & \frac{1}{\sqrt{6}} & -\frac{1}{\sqrt{3}} \\
        -\frac{1}{\sqrt{2}} & \frac{1}{\sqrt{6}} & -\frac{1}{\sqrt{3}} \\
        0 & \frac{2}{\sqrt{6}} & \frac{1}{\sqrt{3}}
    \end{pmatrix}\,.
    \label{eq:6_tetrahedral_frames_2}
\end{align}

Figure~\ref{fig:tetrahedral_arrangements} shows three inequivalent spatial configurations for these spin orientations. All of them have the same chirality per plaquette $\chi^{(0)} = \mathbf{S}_i \cdot \left(\mathbf{S}_j \times \mathbf{S}_k\right) = - \frac{4S^3}{3\sqrt{3}}$ for anticlockwise order, and the resulting classical ground state energy gives
\begin{equation}
    H_0 = - \left(J_1 + J_2 + \frac{8S}{3\sqrt{3}} J_\chi \right) N S^2\,,
\end{equation}
where $N$ is the number of sites in the lattice.
As the tetrahedral spin configuration breaks down the sixfold rotation symmetry of the lattice to a symmetry under rotations by $\pi$, the magnon dispersions for the three configurations in Fig.~\ref{fig:tetrahedral_arrangements} have a related symmetry in momentum space where two opposite $M$ points in the Brillouin zone exhibit a gap while the other four host gapless Goldstone modes. The corresponding magnon dispersion is plotted in Fig.~\ref{fig:dispersion_tetrahedral}. A Mathematica script implementing the LSWT calculations is available in the Zenodo repository~\cite{zenodo2026}.

\appsection{The dimer correlation function}

Eventually, we want to compute the connected dimer-dimer correlations
\begin{align}
    C_D^{\mathrm{conn.}}(\vr, \vr')
    &= \frac{1}{N_{\delta}^2} \sum_{\vd, \vd'} C^{\mathrm{conn.}}_{\mathrm{bonds}}(\vr, \vd; \vr', \vd')\,,
\end{align}
where $N_\delta$ is the coordination number of the lattice and 
\begin{equation}
    C^{\mathrm{conn.}}_{\mathrm{bonds}}(\vr, \vd; \vr', \vd') = \braket{\left(\hat{\mathbf{S}}_{\vr} \cdot \hat{\mathbf{S}}_{\vr + \vd}\right) \left(\hat{\mathbf{S}}_{\vr'} \cdot \hat{\mathbf{S}}_{\vr' + \vd'}\right)} - \braket{\hat{\mathbf{S}}_{\vr} \cdot \hat{\mathbf{S}}_{\vr + \vd}} \braket{\hat{\mathbf{S}}_{\vr'} \cdot \hat{\mathbf{S}}_{\vr' + \vd'}}
\end{equation}
denotes the connected correlation on bonds $(\vr, \vr + \vd)$ and $(\vr', \vr' + \vd')$.
After performing a rotation into the local frames, expressed as $\hat{\mathbf{S}}_\vr = R_{L(\vr)} \locSb_\vr$ and with the resulting matrix $M^{(\vr, \vd)}:= R_{L(\vr)}^T R_{L(\vr + \vd)}^{\phantom{T}}$ ($L(\vr)$ denoting the sublattice index of site $\vr$), we can write the connected four-spin correlation as
% \begin{widetext}
\begin{align}
C^{\mathrm{conn.}}_{\mathrm{bonds}}(\vr, \vd; \vr', \vd')/S^2
&= M_{13}^{(\vr, \vd)} M_{13}^{(\vr', \vd')} \braket{\locS_\vr^x \locS_{\vr'}^x} 
+ M_{31}^{(\vr,\vd)}M_{31}^{(\vr',\vd')} \braket{\locS_{\vr + \vd}^x S_{\vr'+\vd'}^x} \nonumber \\
&+ M_{13}^{(\vr,\vd)} M_{31}^{(\vr',\vd')} \braket{\locS_\vr^x \locS_{\vr' + \vd'}^x}
+ M_{31}^{(\vr,\vd)} M_{13}^{(\vr',\vd')} \braket{\locS_{\vr + \vd}^x \locS_{\vr'}^x} \nonumber \\
&+ M_{23}^{(\vr,\vd)} M_{23}^{(\vr',\vd')} \braket{\locS_\vr^y \locS_{\vr'}^y} 
+ M_{32}^{(\vr,\vd)}M_{32}^{(\vr',\vd')} \braket{\locS_{\vr + \vd}^y S_{\vr'+\vd'}^y} \nonumber \\
&+ M_{23}^{(\vr,\vd)} M_{32}^{(\vr',\vd')} \braket{\locS_\vr^y \locS_{\vr' + \vd'}^y}
+ M_{32}^{(\vr,\vd)} M_{23}^{(\vr',\vd')} \braket{\locS_{\vr + \vd}^y \locS_{\vr'}^y} \nonumber \\
&+ M_{13}^{(\vr,\vd)} M_{23}^{(\vr',\vd')} \braket{\locS_\vr^x \locS_{\vr'}^y}
+ M_{23}^{(\vr,\vd)} M_{13}^{(\vr',\vd')} \braket{\locS_\vr^y \locS_{\vr'}^x} \nonumber \\
&+ M_{31}^{(\vr,\vd)} M_{32}^{(\vr',\vd')} \braket{\locS_{\vr + \vd}^x \locS_{\vr' + \vd'}^y}
+ M_{32}^{(\vr,\vd)} M_{31}^{(\vr',\vd')} \braket{\locS_{\vr + \vd}^y \locS_{\vr' + \vd'}^x} \nonumber \\
&+ M_{13}^{{\vr, \vd}} M_{32}^{(\vr',\vd')} \braket{\locS_\vr^x \locS_{\vr' + \vd'}^y}
+ M_{32}^{(\vr,\vd)} M_{13}^{(\vr',\vd')} \braket{\locS_{\vr + \vd}^y \locS_{\vr'}^x} \nonumber \\
&+ M_{23}^{(\vr,\vd)} M_{31}^{(\vr',\vd')} \braket{\locS_\vr^y \locS_{\vr' + \vd'}^x}
+ M_{31}^{(\vr,\vd)} M_{23}^{(\vr',\vd')} \braket{\locS_{\vr + \vd}^x \locS_{\vr'}^y} + \mathcal{O}(1)\,.
\label{eq:6_dimer_dimer_expression}
\end{align}
The expectation values evaluate to
\begin{align}
    \braket{\locS_\vr^x \locS_{\vr'}^x} &= \frac{S}{2N} \sumprime_\kb e^{\ui \kb \cdot (\vr - \vr')} \left|u_\kb + v_\kb\right|^2\,, \\
    \braket{\locS_\vr^y \locS_{\vr'}^y} &= \frac{S}{2N} \sumprime_\kb e^{\ui \kb \cdot (\vr - \vr')} \left|u_\kb - v_\kb\right|^2\,, \\
    \braket{\locS_\vr^x \locS_{\vr'}^y} &= \frac{1}{\ui} \frac{S}{2N} \sumprime_\kb e^{\ui \kb \cdot (\vr - \vr')} \left(u_\kb^{\phantom{*}} v_\kb^* - v_\kb^{\phantom{*}} u_\kb^* - 1\right)\,, \label{eq:6_correlator_xy}\\
    \braket{\locS_\vr^y \locS_{\vr'}^x} &= \braket{\locS_{\vr'}^x \locS_{\vr}^y}^* 
\end{align}
To arrive at Eq.~\eqref{eq:6_correlator_xy}, we applied the identity $|u_\kb|^2 - |v_\kb|^2 = 1$. The terms mixing $\locS^x$ or $\locS^y$ with $\locS^z$ vanish at leading order as they contain an odd number of bosonic operators. Finally using Wick's theorem, the $zz$ correlator becomes
\begin{align}
    \braket{\locS_\vr^z \locS_{\vr'}^z} &= S^2 - S \braket{\hat{n}(\vr)} - S \braket{\hat{n}(\vr')} + \braket{\hat{n}(\vr) \hat{n}(\vr')}\,.
\end{align}
The expectation value of the occupation number operator evaluates as
\begin{equation}
    \braket{\hat{n}(\vr)} = \braket{\cre_\vr \ann_\vr} = \frac{1}{N} \sumprime_\kb |v_\kb|^2\,.
\end{equation}

The density-density correlator can be written down in the form
\begin{equation}
    \braket{\hat{n}(\vr) \hat{n}(\vr')} = \braket{\hat{n}}^2 + \braket{\hat{n}(\vr) \hat{n}(\vr')}_c\,,
\end{equation}
where the connected part is 
\begin{align}
    \braket{\hat{n}(\vr) \hat{n}(\vr')}_c &= \frac{1}{N^2} \sumprime_{\kb, \kb'} e^{\ui (\kb - \kb') \cdot (\vr - \vr')} \left(|v_\kb|^2 |u_{\kb'}|^2 + v_\kb^{\phantom{*}} u_\kb^* u_{\kb'}^{\phantom{*}} v_{\kb'}^*\right)\,.
    \label{eq:6_density_density_correlation}
\end{align}
Connected here means that in the Wick expansion, we only keep terms that connect the two different sites $\vr$ and $\vr'$. Note that when this expression Eq.~\eqref{eq:6_density_density_correlation} is evaluated on a computer, the two sums over $\kb$ and $\kb'$ can be separated to reduce the computational complexity from $\mathcal{O}(N_\kb^2)$ to $\mathcal{O}(N_\kb)$.
The time dependence can be introduced trivially via
\begin{equation}
    \annB_\kb(t) = e^{-\ui \varepsilon_\kb t} \,\annB_\kb\,,
\end{equation}
resulting in the introduction of time-dependent exponentials such as in the expressions
\begin{align}
    \braket{\locS_\vr^x(t) \locS_{\vr'}^x(0)} &= \frac{S}{2N} \sumprime_\kb e^{\ui \kb \cdot (\vr - \vr')} e^{-\ui \varepsilon_\kb t} \left|u_\kb + v_\kb\right|^2
\end{align}
and
\begin{align}
    \braket{\hat{n}(t, \vr) \hat{n}(0, \vr')} &= \frac{1}{N^2} \sumprime_{\kb, \kb'} e^{\ui (\kb - \kb') \cdot (\vr - \vr')} e^{-\ui (\varepsilon_\kb + \varepsilon_{\kb'}) t} \left(|v_\kb|^2 |u_{\kb'}|^2 + v_\kb^{\phantom{*}} u_\kb^* u_{\kb'}^{\phantom{*}} v_{\kb'}^*\right)\,. 
    \label{eq:6_time_dependence_5}
\end{align}

\appsection{Higher-order results for the coplanar spin orders}

For the coplanar spin orders, we chose a convention where the spins lie in the $x$-$z$ plane of the global reference frame. Therefore, all matrix elements $M^{(\vr, \vd)}_{nm}$, $M^{(\vr', \vd')}_{nm}$ that involve a $y$-component (i.e., $n=2$ or $m=2$) vanish.
As a result, the formula for the dimer-dimer correlations at order $\mathcal{O}(S^3)$ simplifies drastically:
\begin{align}
C^{\mathrm{conn.}}_D(\vr, \vr') &= \frac{S^2}{N_\delta^2} \sum_{\vd, \vd'} \left[ \braket{\locS_\vr^x \locS_{\vr'}^x} + \braket{\locS_{\vr + \vd}^x \locS_{\vr' + \vd'}^x}  - \braket{\locS_\vr^x \locS_{\vr' + \vd'}^x} - \braket{\locS_{\vr + \vd}^x \locS_{\vr'}^x} \right] 
\cdot \sin\left(\vartheta_{\vr}-\vartheta_{\vr + \vd}\right) \sin\left(\vartheta_{\vr'} - \vartheta_{\vr' + \vd'}\right) + \mathcal{O}(S^2)\,.
\label{eq:connected_bond_corrs_OS2}
\end{align}
Whereas in the case of the $120^\circ$ order, this expression gives indeed the leading-order contribution to the dimer-dimer correlations in spin-wave theory, the angular prefactors $\sin\left(\vartheta_{\vr}-\vartheta_{\vr + \vd}\right) \sin\left(\vartheta_{\vr'} - \vartheta_{\vr' + \vd'}\right)$ vanish for the stripe order.
Consequently, the leading-order contribution to the dimer-dimer correlations in the stripe phase arises from higher-order terms in the Holstein-Primakoff expansion.
Neglecting higher-order terms in the ground-state approximation to the Hamiltonian, we find from the expansion of the operators the following full expression for the connected bond correlations for a coplanar order:
\begin{align}
    C^{\mathrm{conn.}}_{\mathrm{bonds}}(\vr, \vd; \vr',\vd') &= \cos\left(\vartheta_\vr - \vartheta_{\vr + \vd}\right) \cos\left(\vartheta_{\vr'} - \vartheta_{\vr' + \vd'}\right)  \left[\braket{\locS^z_\vr \locS^z_{\vr + \vd} \locS^z_{\vr'} \locS^z_{\vr' + \vd'}}_c + \braket{\locS^x_\vr \locS^x_{\vr + \vd} \locS^x_{\vr'}\locS^x_{\vr'+\vd'}}_c \right] \nonumber \\
    &+ \braket{\locS^y_\vr \locS^y_{\vr + \vd} \locS^y_{\vr'}\locS^y_{\vr'+\vd'}}_c \nonumber \\
    &+ \cos\left(\vartheta_\vr - \vartheta_{\vr + \vd}\right) \left[\braket{\locS^x_\vr \locS^x_{\vr + \vd} \locS^y_{\vr'}\locS^y_{\vr'+\vd'}}_c + \braket{\locS^z_\vr \locS^z_{\vr + \vd} \locS^y_{\vr'}\locS^y_{\vr'+\vd'}}_c \right] \nonumber \\
    &+ \cos\left(\vartheta_{\vr'} - \vartheta_{\vr' + \vd'}\right) \left[\braket{\locS^y_\vr \locS^y_{\vr + \vd} \locS^x_{\vr'}\locS^x_{\vr'+\vd'}}_c + \braket{\locS^y_\vr \locS^y_{\vr + \vd} \locS^z_{\vr'}\locS^z_{\vr'+\vd'}}_c \right]\nonumber \\
    &+ \cos\left(\vartheta_\vr - \vartheta_{\vr + \vd}\right) \cos\left(\vartheta_{\vr'} - \vartheta_{\vr' + \vd'}\right) \left[\braket{\locS^z_\vr \locS^z_{\vr + \vd} \locS^x_{\vr'}\locS^x_{\vr'+\vd'}}_c + \braket{\locS^x_\vr \locS^x_{\vr + \vd} \locS^z_{\vr'}\locS^z_{\vr'+\vd'}}_c \right] \nonumber \\
    &+ \sin\left(\vartheta_\vr - \vartheta_{\vr + \vd}\right) \sin\left(\vartheta_{\vr'} - \vartheta_{\vr' + \vd'}\right) \left[\braket{\locS^z_\vr \locS^x_{\vr + \vd} \locS^z_{\vr'}\locS^x_{\vr'+\vd'}}_c + \braket{\locS^x_\vr \locS^z_{\vr + \vd} \locS^x_{\vr'}\locS^z_{\vr'+\vd'}}_c \right.\nonumber \\
    &\left. - \braket{\locS^z_\vr \locS^x_{\vr + \vd} \locS^x_{\vr'}\locS^z_{\vr'+\vd'}}_c - \braket{\locS^x_\vr \locS^z_{\vr + \vd} \locS^z_{\vr'}\locS^x_{\vr'+\vd'}}_c \right]\,.
\end{align}

The subscript $\braket{\ldots}_c$ here denotes that we only consider the connected contributions to the expectation value that involves Wick contractions that connect the two dimer pairs $(\vr, \vr + \vd)$ and $(\vr', \vr' + \vd')$.
We give here the explicit result at order $\mathcal{O}(S^2)$. It reads:

\begin{align}
    C^{\mathrm{conn.}}_{\mathrm{bonds}}(\vr, \vd; \vr',\vd') &= \cos\left(\vartheta_\vr - \vartheta_{\vr + \vd}\right) \cos\left(\vartheta_{\vr'} - \vartheta_{\vr' + \vd'}\right)\Big\{ S^2 \left[\braket{\hat{n}(\vr) \hat{n}(\vr')}_c + \braket{\hat{n}(\vr)\hat{n}(\vr'+\vd')}_c \right. \nonumber \\
    &\left. \qquad \qquad + \braket{\hat{n}(\vr + \vd) \hat{n}(\vr')}_c + \braket{\hat{n}(\vr + \vd) \hat{n}(\vr' + \vd')}_c \right] \nonumber \\
    & \qquad \qquad + \braket{\locS^x_\vr \locS^x_{\vr'}} \braket{\locS^x_{\vr + \vd} \locS^x_{\vr' + \vd'}} + \braket{\locS^x_\vr \locS^x_{\vr' + \vd'}} \braket{\locS^x_{\vr + \vd} \locS^x_{\vr'}} \Big\} \nonumber \\
    & + \braket{\locS^y_\vr \locS^y_{\vr'}} \braket{\locS^y_{\vr + \vd} \locS^y_{\vr' + \vd'}} + \braket{\locS^y_\vr \locS^y_{\vr' + \vd'}} \braket{\locS^y_{\vr + \vd} \locS^y_{\vr'}} \nonumber \\
    & + \cos\left(\vartheta_\vr - \vartheta_{\vr + \vd}\right) \Big\{ \braket{\locS^x_\vr \locS^y_{\vr'}} \braket{\locS^x_{\vr + \vd} \locS^y_{\vr' + \vd'}} + \braket{\locS^x_\vr \locS^y_{\vr' + \vd'}} \braket{\locS^x_{\vr + \vd} \locS^y_{\vr'}} \nonumber \\
    & \qquad \qquad - S \left[\braket{\hat{n}(\vr) \locS^y_{\vr'} \locS^y_{\vr' + \vd'}}_c + \braket{\hat{n}(\vr + \vd) \locS^y_{\vr'} \locS^y_{\vr' + \vd'}}_c \right] \Big\} \nonumber \\
    & + \cos\left(\vartheta_{\vr'} - \vartheta_{\vr' + \vd'}\right) \Big\{ \braket{\locS^y_\vr \locS^x_{\vr'}} \braket{\locS^y_{\vr + \vd} \locS^x_{\vr' + \vd'}} + \braket{\locS^y_\vr \locS^x_{\vr' + \vd'}} \braket{\locS^y_{\vr + \vd} \locS^x_{\vr'}} \nonumber \\
    & \qquad \qquad - S \left[\braket{\locS^y_{\vr} \locS^y_{\vr+\vd} \hat{n}(\vr')}_c+ \braket{\locS^y_{\vr} \locS^y_{\vr + \vd}\hat{n}(\vr' + \vd')}_c\right] \Big\} \nonumber \\
    & - \cos\left(\vartheta_\vr - \vartheta_{\vr + \vd}\right) \cos\left(\vartheta_{\vr'} - \vartheta_{\vr' + \vd'}\right) S \Big\{ \braket{\hat{n}(\vr) \locS^x_{\vr'} \locS^x_{\vr' + \vd'}}_c + \braket{\hat{n}(\vr + \vd) \locS^x_{\vr'} \locS^x_{\vr' + \vd'}}_c \nonumber \\
    & \qquad \qquad + \braket{\locS^x_{\vr} \locS^x_{\vr + \vd} \hat{n}(\vr')}_c + \braket{\locS^x_{\vr} \locS^x_{\vr + \vd} \hat{n}(\vr' + \vd')}_c \Big\} \nonumber \\
    & + \sin\left(\vartheta_\vr - \vartheta_{\vr + \vd}\right) \sin\left(\vartheta_{\vr'} - \vartheta_{\vr' + \vd'}\right) \Big\{ \left[S^2-2S\braket{\hat{n}}\right] \nonumber \\
    &\qquad \qquad \cdot \left[\braket{\locS^x_{\vr + \vd} \locS^x_{\vr' + \vd'}} + \braket{\locS^x_{\vr} \locS^x_{\vr'}} - \braket{\locS^x_{\vr + \vd} \locS^x_{\vr'}} - \braket{\locS^x_{\vr} \locS^x_{\vr' + \vd'}} \right] \Big\} \nonumber \\
    & \qquad \qquad - S \left[\braket{\hat{n}(\vr) \locS^x_{\vr + \vd} \locS^x_{\vr' + \vd'}}_c + \braket{\locS^x_{\vr + \vd} \hat{n}(\vr') \locS^x_{\vr'+\vd'}}_c \right. \nonumber \\
    & \qquad \qquad + \braket{\locS^x_{\vr} \hat{n}(\vr + \vd) \locS^x_{\vr'}}_c + \braket{\locS^x_\vr \locS^x_{\vr'} \hat{n}(\vr' + \vd')}_c \nonumber \\
    & \qquad \qquad - \braket{\hat{n}(\vr) \locS^x_{\vr + \vd} \locS^x_{\vr'}}_c - \braket{\locS^x_{\vr + \vd} \locS^x_{\vr'} \hat{n}(\vr' + \vd')}_c \nonumber \\
    & \qquad \qquad \left. - \braket{\locS^x_{\vr} \hat{n}(\vr + \vd) \locS^x_{\vr' + \vd'}}_c - \braket{\locS^x_{\vr}  \hat{n}(\vr') \locS^x_{\vr' + \vd'}}_c \right] \Big\} + \mathcal{O}(S)\,.
    \label{eq:6_dimer_corrs_higher_order}
\end{align}
These terms can be readily implemented and evaluated numerically.

We find for instance the following explicit correlation
\begin{align}
\braket{\hat{n}(t,\vr) \locS^y_{\vr'} \locS^y_{\vr' + \vd'}}_c &= \frac{S}{2N^2} \sumprime_{\kb, \kb'} e^{-\ui (\varepsilon_\kb + \varepsilon_{\kb'})t}
e^{\ui \kb \cdot (\vr - \vr')} e^{\ui \kb' \cdot (\vr - \vr' - \vd')} \nonumber \\
& \quad \cdot \left[ \left(u_\kb^2 -u_\kb v_\kb\right) \left(v_{\kb'}^2 - u_{\kb'}v_{\kb'}\right) + \left(v_{\kb}^2-u_{\kb}v_{\kb}\right) \left(u_{\kb'}^2 - u_{\kb'}v_{\kb'}\right) \right]\,,
\end{align}
in contrast to
\begin{align}
\braket{\hat{n}(t,\vr) \locS^x_{\vr + \vd}(t) \locS^x_{\vr'+\vd'}}_c &= \frac{S}{2N^2} \sumprime_{\kb, \kb'} e^{-\ui \varepsilon_{\kb'} t}
e^{\ui \kb \cdot (-\vd)} e^{\ui \kb' \cdot (\vr - \vr' - \vd')} \nonumber \\
& \quad \cdot \left[ \left(u_\kb^2 +u_\kb v_\kb\right) \left(v_{\kb'}^2 +u_{\kb'}v_{\kb'}\right) + \left(v_{\kb}^2+u_{\kb}v_{\kb}\right) \left(u_{\kb'}^2 + u_{\kb'}v_{\kb'}\right) \right]\,.
\end{align}
Note that both the energy $\varepsilon_\kb$ and the Bogoliubov coefficients $u_\kb$, $v_\kb$ are even functions in the momentum argument. The latter coefficients have been chosen real-valued for the coplanar spin order, hence we drop the complex conjugation in the expressions above.
These considerations conclude our discussion of the semiclassical analytics as all formulas used for the evaluation of the dynamical dimer structure factor in spin-wave theory have been derived.
\end{widetext}

\end{document}